\documentclass[pra,twocolumn,superscriptaddress]{revtex4}

\usepackage{graphicx}
\usepackage{amssymb}
\usepackage{times}
\usepackage{amsmath}

\begin{document}

\title{Conditional quantum-state engineering 
using ancillary squeezed-vacuum states
}

\author{Hyunseok~Jeong} \affiliation{Department of Physics, University of Queensland, St Lucia,
Queensland 4072, Australia}

\author{Andrew~M.~Lance} \affiliation{Quantum Optics Group, Department
of Physics, Faculty of Science, Australian National University,
ACT 0200, Australia}

\author{Nicolai~B.~Grosse} \affiliation{Quantum Optics Group, Department
of Physics, Faculty of Science, Australian National University,
ACT 0200, Australia}

\author{Thomas~Symul} \affiliation{Quantum Optics Group, Department
of Physics, Faculty of Science, Australian National University,
ACT 0200, Australia}

\author{Ping~Koy~Lam} \affiliation{Quantum Optics Group, Department
of Physics, Faculty of Science, Australian National University,
ACT 0200, Australia}

\author{Timothy~C.~Ralph} \affiliation{Department of Physics, University of Queensland, St Lucia, 
Queensland 4072, Australia}

\date{\today}

\begin{abstract}
We investigate an 
optical scheme to conditionally engineer
quantum states  using
a beam splitter, homodyne detection 
and a squeezed vacuum
as an ancillar state. 
This scheme is efficient in producing non-Gaussian
quantum states such as squeezed single photons and
superpositions of coherent states (SCSs).
We show that a SCS with well defined parity and 
high fidelity can be generated from a Fock state of $n\leq4$,
and conjecture that this can be generalized for an arbitrary $n$ Fock state.
We describe our experimental demonstration of this   
scheme using coherent input states and measuring 
experimental fidelities that
are only achievable using quantum resources. 
\end{abstract}


\maketitle

\section{Introduction}

Quantum state engineering 
using measurement induced conditional evolution is an 
important and useful technique 
in the field of quantum optics and quantum information processing~\cite{NIE00}.
It is known that a near deterministic, 
universal set of unitary transformations can be achieved
for qubit systems using
this principle \cite{KNI01,OBR03},
and that arbitrary optical states can 
be engineered conditionally based on discrete 
single photon measurements~\cite{DAK99}. 
Recently, there has been increased interest in conditional 
evolution based on continuous-variable 
measurements~\cite{Paris,Lau03, Bab05, RAL05}.  
In these schemes a quantum system is interacted 
with a prepared ancilla, which is measured via 
a {\it continuous} observable, e.g. a quadrature
variable of the electromagnetic field.  
This has been experimentally demonstrated for 
a system using a beam splitter as the interaction 
and a vacuum state as the ancilla with conditioning 
based on homodyne detection to remotely prepare a
qubit state~\cite{Bab05}. 
A similar system using conditioning 
of adaptive phase measurements has  also been 
discussed~\cite{RAL05}.

Recently, an optical scheme was suggested to
engineer 
interesting continuous-variable 
non-Gaussian quantum states based on a beam-splitter interaction, 
using an ancilla squeezed vacuum state and conditioning homodyne detection
\cite{Lance06}.
It is a difficult task to generate and engineer non-Gaussian 
continuous-variable quantum states in optical fields.
For example, a superposition of free-traveling coherent states (SCS) 
is very hard to generate 
in spite of its potential usefulness 
for quantum information processing
\cite{catapply} and 
fundamental interest in relation to 
Schr\"odinger's cat paradox \cite{Schr}.
As another example, it is highly nontrivial to directly
squeeze a single photon in a squeezing apparatus,
even though the squeezed single photon can also be useful
for quantum information processing applications \cite{Lund04,Wenger,Neer,Ol,Suzuki}.
The scheme in Ref.~\cite{Lance06}
enables one to conditionally squeeze a single photon 
with an arbitrarily high fidelity
using the squeezed vacuum of any finite degree of squeezing.  
It also enables one to transform a two-photon state into
a SCS with an extremely high fidelity.
The principles of the postselection scheme 
were experimentally demonstrated
using coherent states, and 
experimental fidelities were measured that
are only achievable using quantum resources~\cite{Lance06}.

In this paper,
we develop the scheme
presented in Ref.~\cite{Lance06} and fully describe
its experimental demonstration.
In particular, we find that a SCS with well defined parity and 
high fidelity can be generated from a Fock state of $n\leq4$,
and conjecture that this can be generalized for an arbitrary $n$.
We also compare the postselection scheme \cite{Lance06}
with another scheme based on the feedforward method using
the displacement operation \cite{Fil05}.
It is known that the squeezing operation for an arbitrary input state
can be approximately performed
using the highly squeezed vacuum as an ancillar, a beam splitter and
the feedforward method with the displacement operation
\cite{Fil05}.
However, the interesting 
features of the post-selection scheme that we described above
for {\it non-Gaussian} inputs {\it cannot} be achieved using
the feed-forward method in Ref.~\cite{Fil05}.
This shows 
that quantum state engineering using post-selection
is an interesting tool for quantum 
information processing.
On the other hand, 
for Gaussian coherent states,
the standard feedforward scheme can be {\it modified} to perform
the same transformation with the postselection scheme.
In this case, the postselection scheme can be understood as
an alternative method to the feedforward scheme \cite{Fil05,JF05}.

\section{Engineering various input states
 using a squeezed vacuum and postselection}

\begin{figure}[ht!]
\begin{center}
\includegraphics[width=\columnwidth]{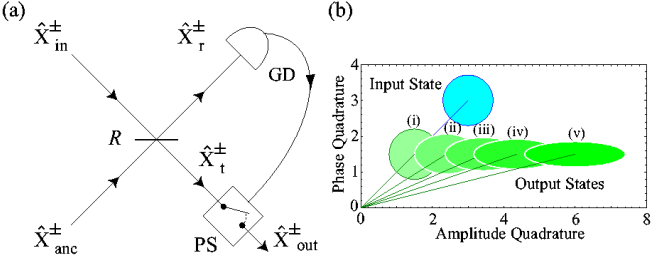}
\caption{(a) Schematic of the post-selection protocol. 
$\hat{X}_{\rm in}^{\pm}$: amplitude (+) and phase (-) 
quadratures of the input state; (anc) 
ancilla squeezed vacuum 
state; (r) reflected; (t) transmitted;  (out)
post-selected output state. 
$R$: beam-splitter reflectivity; GD: gate detector;
PS: post-selection protocol.
(b) Standard deviation contours of the 
Wigner functions of an 
input coherent state (blue) and post-selected 
output states (green) for 
$R=0.75$ and varying ancilla state 
squeezing of (i) $s=0$,
(ii) $s=0.35$, (iii) $s=0.69$, 
(iv) $s=1.03$ and (v) ideal squeezing. 
}\label{ExptSetup0}
\end{center}
\end{figure}

Our conditional transformation scheme is depicted in Fig.~\ref{ExptSetup0}.
The squeezed vacuum used as the ancilla state in our scheme
is represented as ${\hat S}(s)|0\rangle$
with the squeezing operator ${\hat S}(s) = {\rm exp}[-(s/2)(\hat{a}^2 - \hat{a}^{\dagger 2})]$,
where $s$ is the squeezing parameter and $\hat a$ is the annihilation operator.
The Wigner function of the squeezed vacuum is 
\begin{equation}
W_{\rm sqz}(\alpha;s)=\frac{2}{\pi}\exp[-2(\alpha^{+})^2e^{-2s}-2(\alpha^{-})^2e^{2s}],
\end{equation}
where $\alpha=\alpha^{+}+i\alpha^{-}$ with real quadrature variables
$\alpha^{+}$ and $\alpha^{-}$.
The first step of our transformation protocol is to interfere 
the input field with the ancilla state
on a beam-splitter as shown in Fig.~\ref{ExptSetup0}~(a).
The beam-splitter operator ${\hat{B}}$ acting on
 modes $\hat a$ and $\hat b$ is represented as 
\begin{equation}
\hat{B}(\theta)=\exp \{\frac{\theta}{2}
(\hat{a}^{\dagger }\hat{b}
-\hat{b}^{\dagger }\hat{a})\},
\end{equation} 
where the reflectivity is defined as 
$R=\sin^2(\theta /2)$ and where $T=1-R$.
A homodyne measurement is performed on the amplitude
quadrature on the reflected field mode, with the measurement
result denoted as $X^+_{\rm r}$. 
The transmitted state is post-selected for $|X^+_{\rm r}|<x_0$, where
the post-selection threshold $x_0$ is 
determined by the required fidelity
between the output state and the ideal target state.
As we will see later in this section, the postselection process
in our scheme plays a crucial role in engineering non-Gaussian states
with finite squeezing resources.

\subsection{Squeezing a single photon}

We first consider a single-photon state input, $|1\rangle$,
and a squeezed single photon, 
${\hat S}(s^\prime)|1\rangle$, as the target state.
The Wigner function of the single photon state 
is
\begin{equation}
W^{|1\rangle}_{\rm in}(\alpha)=\frac{2}{\pi}\exp[-2|\alpha|^2](4|\alpha|^2-1).
\end{equation}
After interference via 
the beam-splitter, the resulting two-mode state becomes
\begin{equation} 
W^{(1)}(\alpha,\beta)=W^{|1\rangle}_{\rm in}\big( \sqrt{T}\alpha+\sqrt{R}\beta\big)
W_{\rm anc}\big(-\sqrt{R}\alpha+\sqrt{T}\beta\big),
\label{eq:bscom}
\end{equation}
where $W_{\rm anc}= W_{\rm sqz}(\alpha;s)$ and 
$\beta=\beta^{+}+i\beta^{-}$.
Note that the superscript number in the parentheses,
``$(1)$'', indicates that the input state was the single-photon Fock state.
We will use this notation, ``$(n)$'', for various quantities
throughout the paper to denotes that the input state was 
the $n$-photon Fock state.
The transmitted state after the homodyne detection of 
the reflected state is
\begin{equation}
W^{(1)}_{\rm out}(\alpha;X^+_{\rm r})=P_{(1)}(X^+_{\rm r})^{-1}\int_{-\infty}^{\infty}
d\beta^{-}W^{(1)}(\alpha,\beta^{+}=X^+_{\rm r},\beta^{-}),
\end{equation}
where the normalization parameter is 
\begin{eqnarray}
\begin{aligned}
&P_{(1)}(X^+_{\rm r})=\int_{-\infty}^\infty d^2\alpha
d\beta^{-}W^{(1)}(\alpha,\beta^{+}=X^+_{\rm r},\beta^{-})\\
&~~=\frac{2e^{-s-2{X_{\rm r}^+}^2/(R+Te^{2s})}
(T^2e^{4s}+e^{2s}TR+4R{X_{\rm r}^+}^2)
}
{\pi(R+Te^{2s})^2\sqrt{T+e^{-2s}}}.
\end{aligned}
\end{eqnarray}
If the measurement result is $X_{\rm r}^+=0$, 
the Wigner function of the output state becomes 
\begin{widetext}
\begin{equation}
\label{eq:Wout1}
W^{(1)}_{\rm out}(X^+_{\rm r}=0;\alpha)
=\frac{2}{\pi}e^{-2[e^{-2s^\prime}(\alpha^{+})^2+e^{2s^\prime}
(\alpha^{-})^2]}
\big\{4e^{-2s^\prime}(\alpha^{+})^2+4e^{2s^\prime}(\alpha^{-})^2-1\big\}
\equiv W_{\rm ssp}(\alpha), 
\end{equation}
\end{widetext}
where 
\begin{equation}
s^\prime=-\frac{1}{2}\ln [T + e^{-2s}R].
\label{equ:sp1}
\end{equation}
One can immediately notice that the output 
state in Eq.~(\ref{eq:Wout1})
is {\it exactly} the Wigner function
$W_{\rm ssp}(\alpha)$
 of a 
squeezed single photon, $\hat{S}(s^\prime)|1\rangle$.
We note that the output squeezing $s^\prime$ 
can be arbitrarily close to 
the squeezing of the ancilla state $s$ by making $R$ close to zero.

\begin{figure}
\centerline{\scalebox{0.58}{\includegraphics{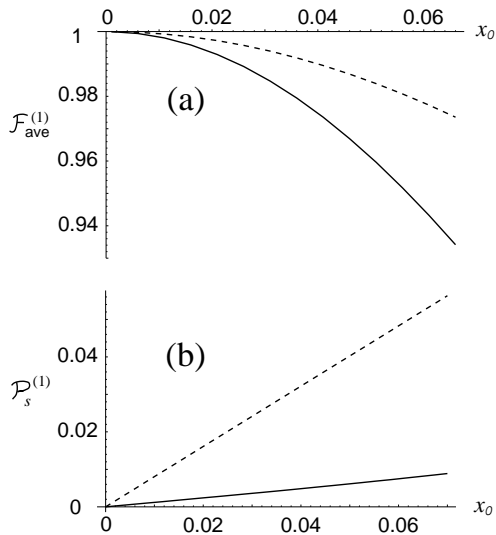}}}
\caption{
(a)The average fidelity ${\cal F}^{(1)}_{\rm ave}$
between the post-selected output state of an 
input single photon state $|1\rangle$
and the ideal squeezed single photon target state 
${\hat S}(s^\prime)|1\rangle$.
(b) The success probability ${\cal P}_{\rm s}^{(1)}$
against the post-selection threshold $x_0$.
Solid line: beam-splitter reflectivity $R=0.98$, ancillary state squeezing 
$s=0.7$ and target state squeezing of $s^\prime=0.67$. Dashed line: beam-splitter 
reflectivity of $R=0.5$, ancillary state squeezing $s=-0.7$ and target 
state squeezing of $s¡¯=-0.464$.
}
\label{fig:sp}
\end{figure}
\begin{figure}[ht!]
\vspace{0.4cm}
\centerline{(a)~~~~~~~~~~~~~~~~~~~~~~~~~~~~~~~~~~~~~~~~~~~~(b)~~~~~~~~~~~~~}
\centerline{\scalebox{0.43}{\includegraphics{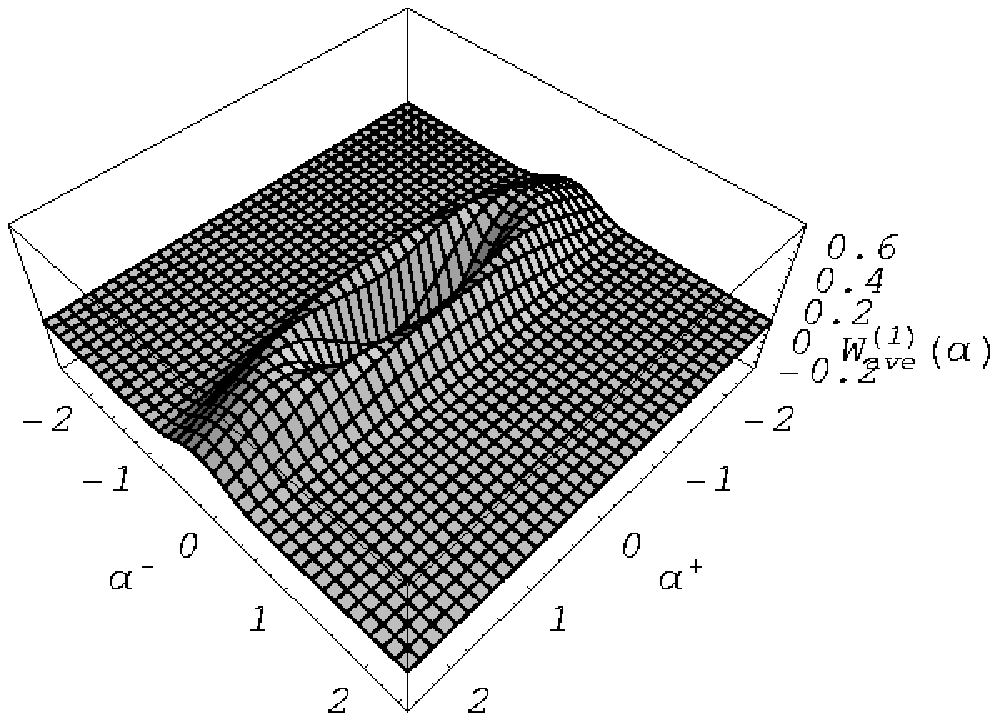}}
\scalebox{0.43}{\includegraphics{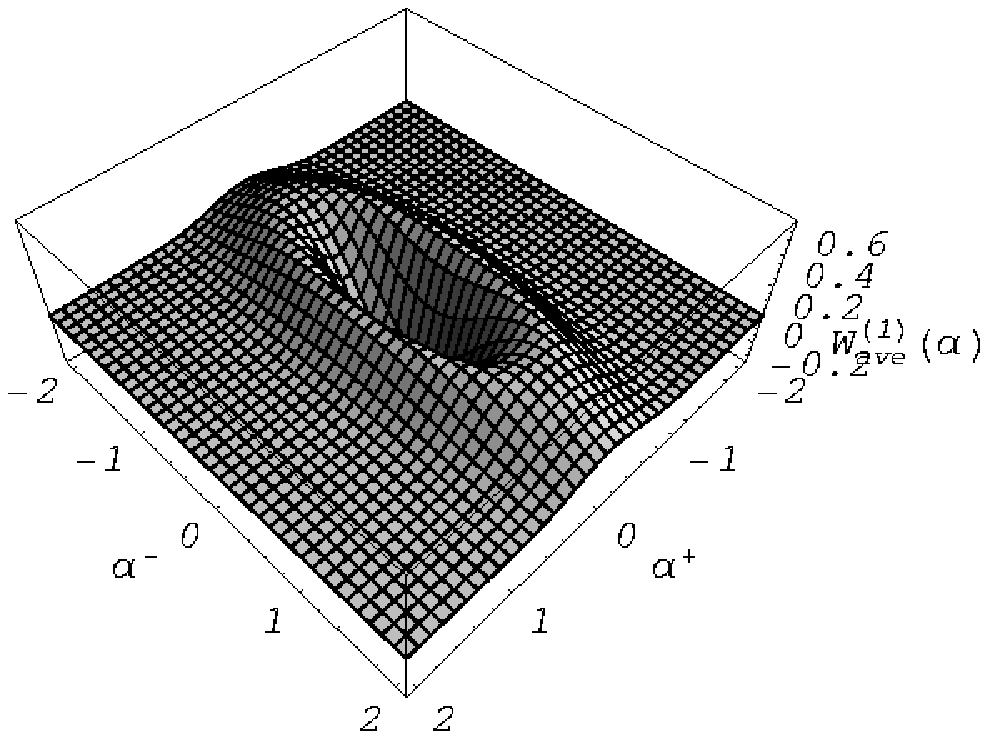}}}
\caption{
The average Wigner function 
$W_{\rm ave}^{(1)}(\alpha)$ of the output state corresponding to
${\cal F}_{\rm ave}^{(1)}=0.99$ (a) for $x_0=0.025$, $R=0.98$, $s=0.7$
and ${\cal P}^{(1)}_{\rm s}=0.003$
and for (b) $x_0=0.04$, $R=1/2$, $s=-0.7$ and ${\cal P}^{(1)}_{\rm s}=0.016$.
}
\label{1photon_Input}
\end{figure}

For a nonzero post-selection threshold criteria 
$|X^+_{\rm r}|<x_0$ the corresponding success probability
is given by 
\begin{equation}
{\cal P}^{(1)}_{\rm s}(x_0)=\int_{-x_0}^{x_0} d X^+_{\rm r} P_{(1)}(X^+_{\rm r})
\end{equation}
and the average Wigner function $W^{(1)}_{\rm ave}(\alpha;x_0)$ is
\begin{equation}
W^{(1)}_{\rm ave}(\alpha;x_0)=\frac{\int_{-x_0}^{x_0} dX^+_{\rm r}
 P_{(1)}(X^+_{\rm r})W_{\rm out}^{(1)}
(\alpha,X^+_{\rm r})}
{\int_{-x_0}^{x_0}d{\tilde X^+_{\rm r}}
P_{(1)}({\tilde X^+_{\rm r}})}.
\end{equation}
The fidelity between the output state (with 
the measurement result $X^+_{\rm r}$)
and the ideal target state is obtained as
\begin{equation} 
{\cal F}_{(1)}(X^+_{\rm r})=\pi\int^\infty_{-\infty}
d^2\alpha W^{(1)}_{\rm out}(\alpha;X^+_{\rm r})W_{\rm ssp}(\alpha).
\end{equation}
The average fidelity of the output state for a postselection threshold $x_0$ is
\begin{equation}
{\cal F}^{(1)}_{\rm ave}(x_0)=
\frac{\int_{-x_0}^{x_0} dX^+_{\rm r} P_{(1)}(X^+_{\rm r}){\cal F}_{(1)}(X^+_{\rm r})}
{\int_{-x_0}^{x_0}d{\tilde X^+_{\rm r}}
P_{(1)}({\tilde X^+_{\rm r}})}.
\label{eq:af}
\end{equation}
Alternatively, the  average fidelity for
a threshold $x_0$ can be obtained using
the average Wigner function as
\begin{equation}
{\cal F}^{(1)}_{\rm ave}(x_0)=\pi\int^\infty_{-\infty}
d^2\alpha W^{(1)}_{\rm ave}(\alpha;x_0)W_{\rm ssp}(\alpha).
\label{eq:af2}
\end{equation}
We use this  average fidelity measure to characterize the efficacy 
of our protocol for nonzero thresholds.

The average fidelity ${\cal F}^{(1)}_{\rm ave}$ and success probability
 ${\cal P}^{(1)}_{\rm s}$
have been plotted for a couple of cases in Fig.~\ref{fig:sp}.
Suppose that the ancillar squeezing is $s=0.7$ (6.08~dB
phase squeezing) with
the beam splitter ratio $R=0.98$ and 
the post-selection threshold is $x_0=0.025$.
Then, the output squeezing is  $s^\prime=0.67$ (5.82~dB) and
the average fidelity 
is ${\cal F}^{(1)}_{\rm ave}= 0.99$
with the success probability ${\cal P}^{(1)}_{\rm s}=0.003$.
As another example,
if $s=-0.7$ (6.08~dB amplitude squeezing)
and $x_0=0.04$ with $R=1/2$ are assumed,
the results are
$s^\prime=-0.464$ ($4.03$~dB)
and ${\cal F}^{(1)}_{\rm ave}= 0.99$
with ${\cal P}^{(1)}_{\rm s}=0.016$.
The average Wigner functions
corresponding to an average fidelity  ${\cal F}^{(1)}_{\rm ave}=0.99$
for these cases are shown in Fig.~\ref{1photon_Input}.

We emphasize that squeezing a single photon
using the squeezed vacuum with such high fidelities 
and high degrees of the output squeezing
{\it cannot} be achieved by the feedforward method \cite{Fil05}
unless the ancillar squeezing becomes extremely high (which is not realistic).
However, our scheme enables one to perform this interesting task
{\it with any finite degree of the ancillar squeezing}.
Fig.~\ref{fig:c123} shows that the output states obtained by
the feedforward method are distorted but the postselection for an appropriate 
threshold results in the desired state.
Fig.~\ref{fig:ng123} explains the role of the post-selection for a non-Gaussian input:
if the homodyne result $X^+$ is far from zero,
the shape of the output state become distorted, i.e.,
the output state loses the non-Gaussian characteristics.
Therefore, it cannot be 
corrected by the feedforward with any electronic gain, and
the postselection is necessary to select non-Gaussian output states. 
In other words, post-selection around $X^{+}_{\rm r}=0$ preserves
the non-Gaussian features of the input state.

\begin{widetext}
\begin{center}
\begin{figure}
\centerline{(a)~~~~~~~~~~~~~~~~~~~~~~~~~~~~~~~~~~~~~~~~~~~~~~~~(b)
~~~~~~~~~~~~~~~~~~~~~~~~~~~~~~~~~~~~~~~~~~~~~~~~(c)~~~~~~~~~~~~}
\centerline{\scalebox{0.42}{\includegraphics{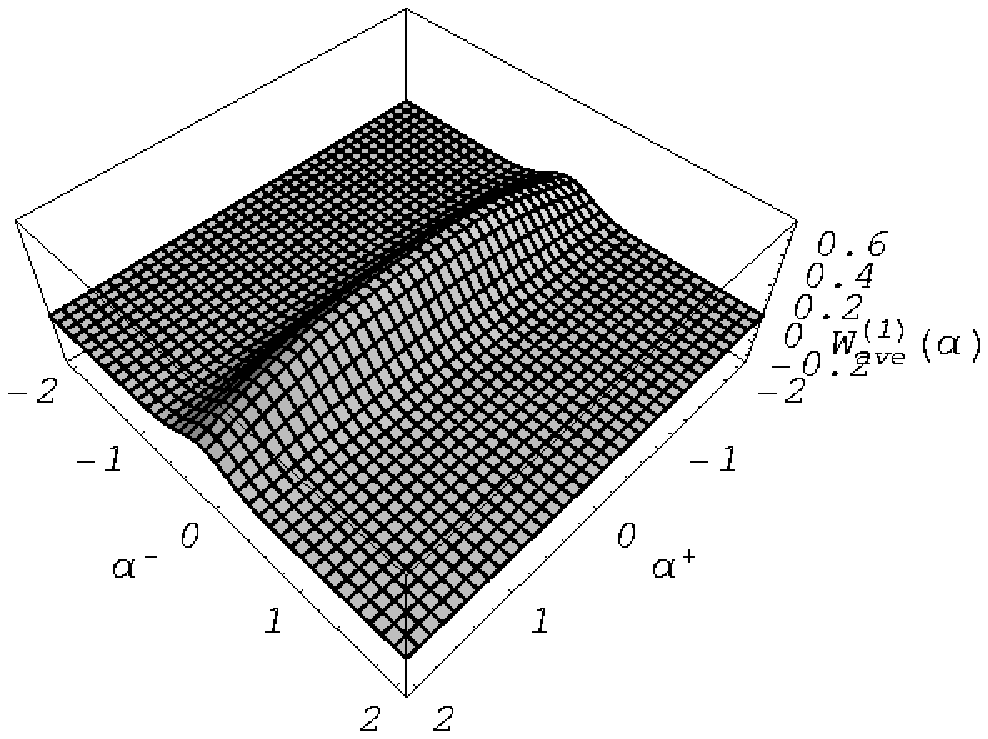}}
\scalebox{0.42}{\includegraphics{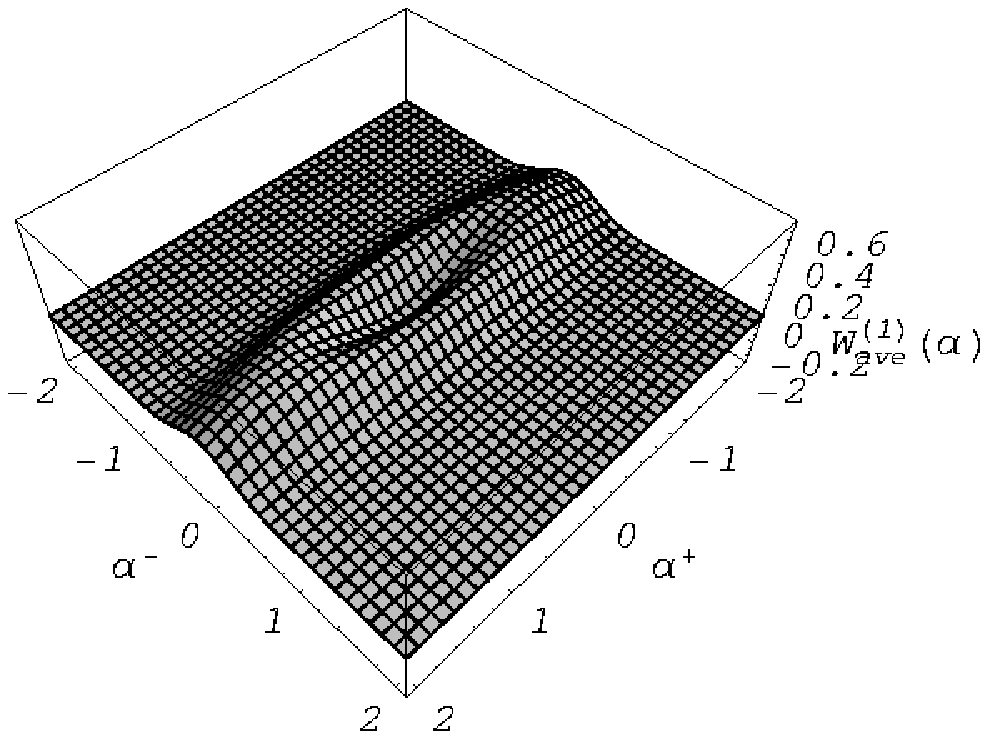}}
\scalebox{0.42}{\includegraphics{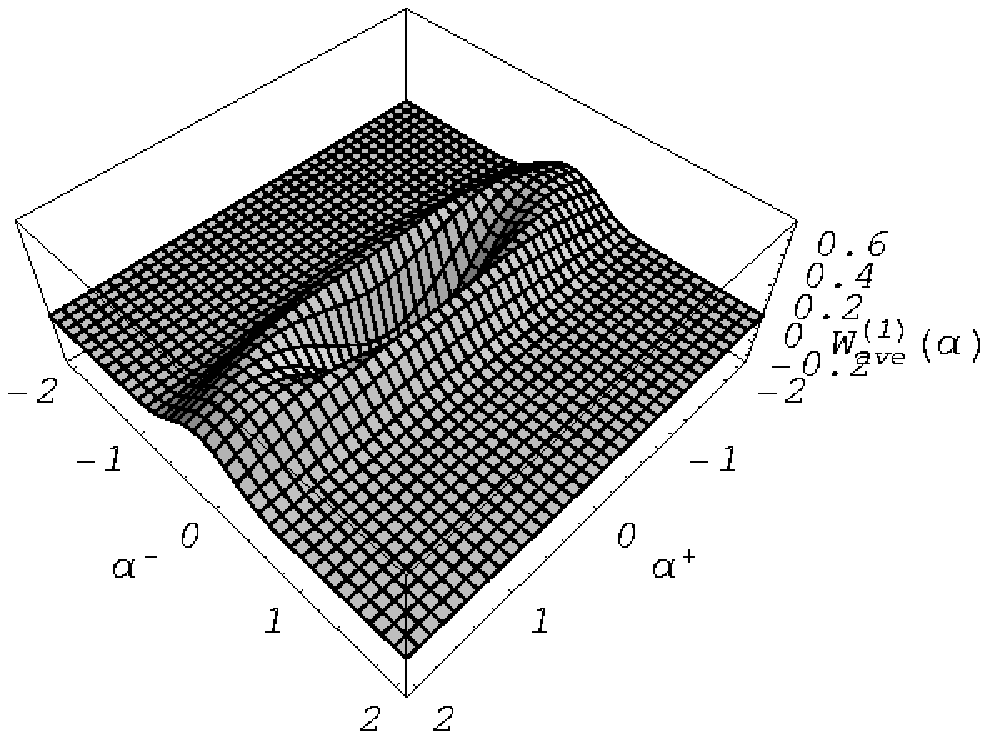}}
}
\caption{
(a) The average Wigner function $W^{(1)}_{\rm ave}(\alpha)$
of the output state obtained from the single-photon input 
with $s=0.7$ and $R=0.98$
using the feedforward scheme with the unity gain.
The non-Gaussian quantum characteristics are washed away. 
(b) If the result obtained by the feedforward method is postselected for $x=0.3$,
the non-Gaussian characteristic begins to emerge. 
(c) 
The average fidelity ${\cal F}_{\rm ave}^{(1)}=0.87$ and the minimum negative
value $W^{(1)}_{\rm ave}(0)\approx-0.48$ are obtained
for threshold $x_0=0.1$.  
}
\label{fig:c123}
\end{figure}
\begin{figure}
\centerline{~~~~~~~~~~~~(a)~~~~~~~~~~~~~~~~~~~~~~~~~~~~~~~~~~~~~~~~~~~~~~~~~~(b)
~~~~~~~~~~~~~~~~~~~~~~~~~~~~~~~~~~~~~~~~~~~~~~~~~~(c)~~~~~~}
\centerline{\scalebox{0.42}{\includegraphics{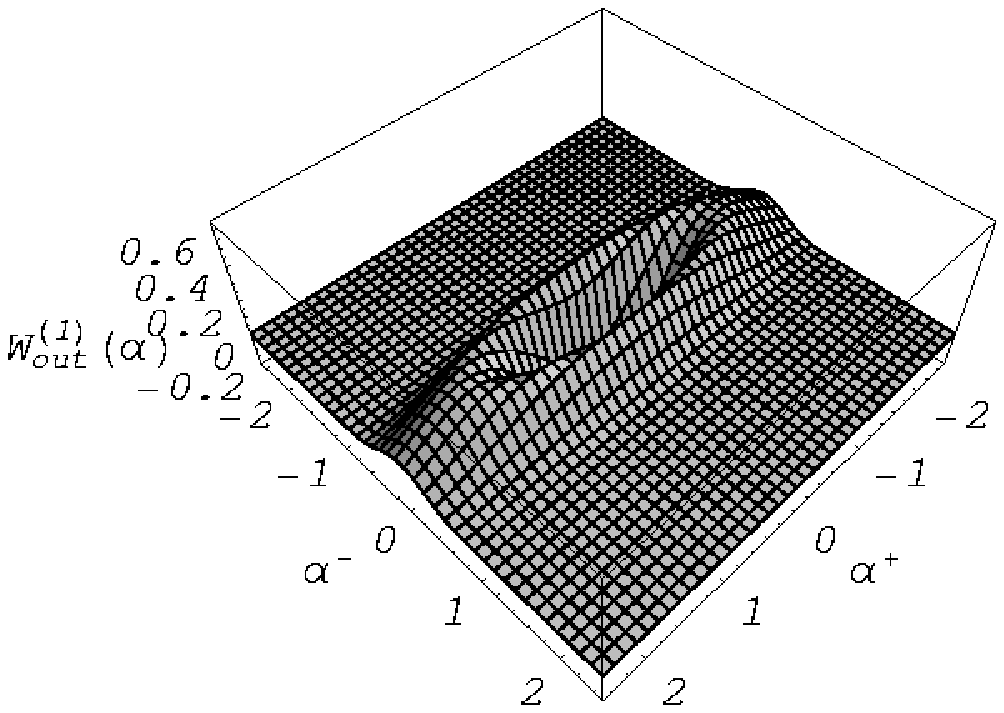}}
\scalebox{0.42}{\includegraphics{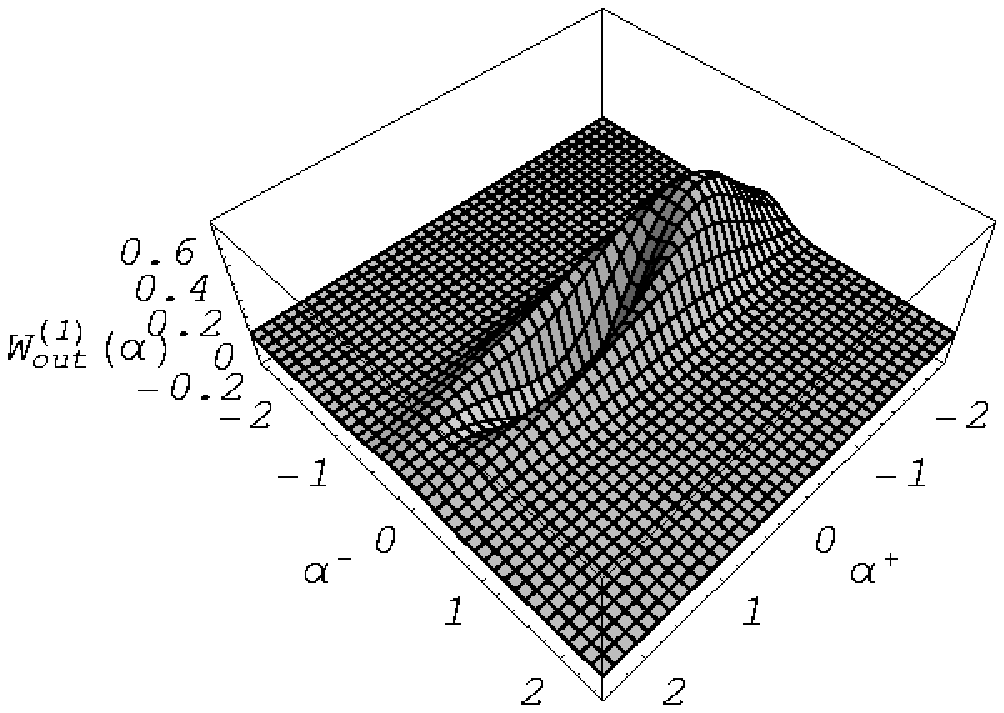}}
\scalebox{0.42}{\includegraphics{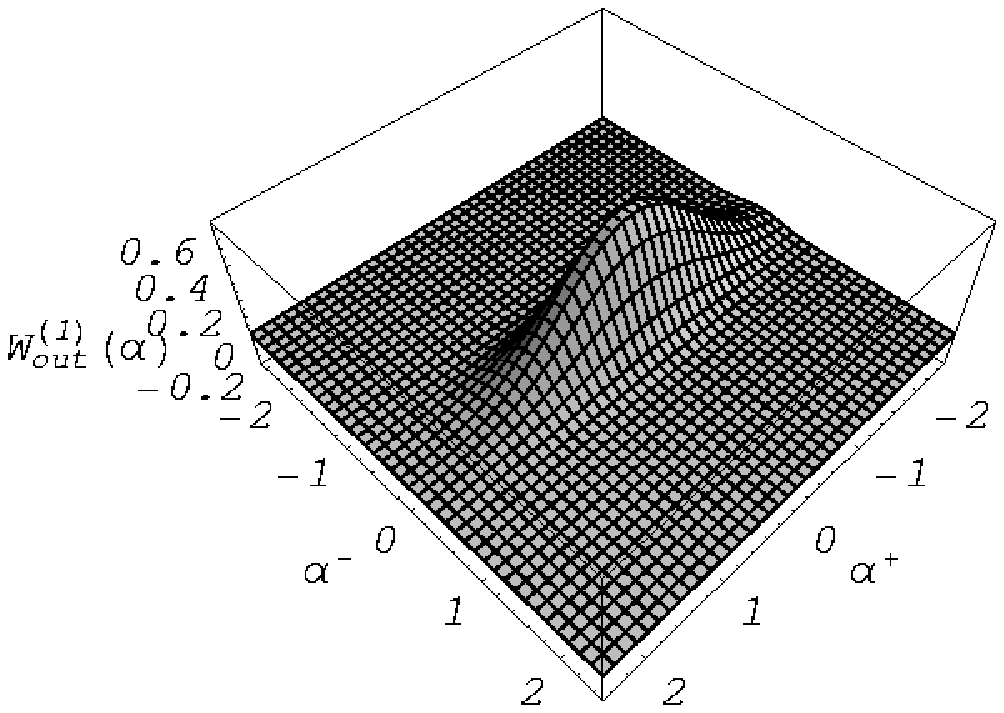}}
}
\caption{
The Wigner function $W^{(1)}_{\rm out}(\alpha)$
of the output state obtained from the single-photon input 
with $s=0.7$ and $R=0.98$. 
(a) If the measurement result is $X^+_{\rm r}=0$,
an ideal squeezed single photon with $s^\prime=0.67$ is  obtained, i.e.,
 the fidelity is ${\cal F}_{(1)}(X_{\rm r}^+=0)=1$.
 However, the non-Gaussian shape is distorted as the measurement results deviate
from zero as (b) ${\cal F}_{(1)}(X_{\rm r}^+=-0.1)=0.67$ and
(c)  ${\cal F}_{(1)}(X_{\rm r}^+=-0.5)=0.03$.
The non-Gaussian characteristics cannot be recovered
by the classical feedforward methods after such distortions.
It is clear from this figure that the postselection is necessary to
preserve the non-Gaussian feature for the output state. 
}
\label{fig:ng123}
\end{figure}
\end{center}
\end{widetext}

\subsection{Converting Fock states to superposition of coherent states}

We now examine Fock states for $n\geq2$ as input states.
Our target states are 
the SCSs
\begin{equation}
|{\rm SCS}_\pm\rangle=\frac{1}
{\sqrt{2\pm 2e^{-2|\gamma|^2}}}(|\gamma\rangle\pm|-\gamma\rangle),
\end{equation}
where $|\gamma\rangle$ is a coherent state of amplitude 
$\gamma=\gamma^{+}+i\gamma^{-}$.
The SCSs are often referred to as ``Schr\"odinger cat states''
due to their characteristics as superpositions of macroscopically
distinguishable states. 
The SCSs $|{\rm SCS}_+\rangle$ and $|{\rm SCS}_-\rangle$
are called even and odd SCSs, respectively.
The even SCS contains only even number of photons
and becomes the vacuum as the amplitude goes to zero, i.e., 
$\alpha\rightarrow0$.
On the other hand, the odd SCS contains only odd number of photons
and approaches the single photon state $|1\rangle$ as the amplitude goes to zero.
The Wigner representations of the even and odd SCSs are
\begin{eqnarray}
W^{\pm}_{\rm scs}(\alpha)=N_\pm\Big\{e^{-2|\alpha-\gamma|^2}
+e^{-2|\alpha+\gamma|^2}~~~~~~~~~~~~~~~~~~~~~~~~\nonumber\\
\pm e^{-2|\gamma|^2}(e^{-2(\alpha+\gamma)^*(\alpha-\gamma)}
+e^{-2(\alpha+\gamma)(\alpha-\gamma)^*})
\Big\},
\label{eq:scs}
\end{eqnarray}
where 
$N_\pm=\{\pi(1 \pm e^{-2|\gamma|^2})\}^{-1}$.
In this subsection, we shall assume $\gamma$ to be {\it pure imaginary}
for simplicity of equations without losing generality.

The output Wigner function, average fidelity
and success probability can be obtained 
in the same way described in Eqs.~(\ref{eq:bscom}) to (\ref{eq:af2}).
The Wigner function
of the output state for the two-photon input with 
the measurement result $X^+_{\rm r}$ and
the beam-splitter reflectivity $R=1/2$ is given by
\begin{eqnarray}
&&W^{(2)}_{\rm out}(\alpha;X^+_{\rm r})= N_2 e^{-G}
\Big\{
1+2Z+(3+16\alpha_i^2)Z^2 
\nonumber\\
&&+(4-16\alpha_i^2)Z^3+(2-32\alpha_i^2+64\alpha_i^4)Z^4
+4{\alpha^\prime_r}^4(1+Z)^4  \nonumber\\
&&-4{\alpha^\prime_r}^2(1+Z)^2(1+3Z+(2-8\alpha_i^2)Z)    
\Big\} 
\label{eq:out2p}
\end{eqnarray}
where $G={2\alpha_i^2+{\alpha_r^\prime}^2
+Z^{-1}(\alpha_r-X^+)^2
+2\alpha_i^2\tanh[s]}$,
${\alpha_r^\prime}=\alpha_r+X^+$,
$Z=e^{2s}$, and $N_2$ is the normalization factor.
The success probability is
${\cal P}^{(2)}_{\rm s}(x_0)=\int_{-x_0}^{x_0} d X^+_{\rm r} P_{(2)}(X^+_{\rm r})$
with
\begin{equation}
P_{(2)}(X^+_{\rm r})=\frac
{J e^{-s-{4{X^+_{\rm r}}^2/K}}}
{\sqrt{\pi(1+e^{-2s})}J^4}
\end{equation}
where 
$J=4e^{6s}+2e^{8s}+(1-8{X^+_{\rm r}}^2)^2
+2e^{2s}(1+8{X^+_{\rm r}}^2)
+e^{4s}(3+32{X^+_{\rm r}}^2)$
and $K=1+e^{2 s}$.
The fidelity is calculated as
${\cal F}_{(2)}(X^+_{\rm r})=\pi\int^\infty_{-\infty}
d^2\alpha W^{(2)}_{\rm out}(\alpha;X^+_{\rm r})W^+_{\rm scs}(\alpha)$.
If the measurement result is $X_{\rm r}^+=0$ with $R=1/2$,
the fidelity between the ideal {\it even} SCS in Eq.~(\ref{eq:scs})
and the output state (\ref{eq:out2p}) is simplified as
\begin{equation}
{\cal F}_{(2)}
(X_{\rm r}^+=0)=\frac{A K^\frac{5}{2}
(1+4e^{2s}+(3-8\gamma^2)e^{4s})^2}
{(1+e^{2 \gamma^2})(1+3e^{2s})^5(1+2e^{4s})}
\label{f1}
\end{equation}
where
$A=4\sqrt{2}\exp[\gamma^2+s+\gamma^2\sinh s/
(2\cosh s)+\sinh s]$.
If the ancillar squeezing is $s=-0.37$ ($3.21$~dB),
the fidelity becomes
${\cal F}_{(2)}(X_{\rm r}^+=0)\approx0.9997$
for  $\gamma=1.1i$, i.e.,
{\it a SCS of amplitude $|\gamma|=1.1$ 
with extremely high fidelity 
can be obtained
from a two-photon Fock state
using the squeezed vacuum of $3.21$~dB squeezing.}

\begin{figure}
\centerline{\scalebox{0.63}{\includegraphics{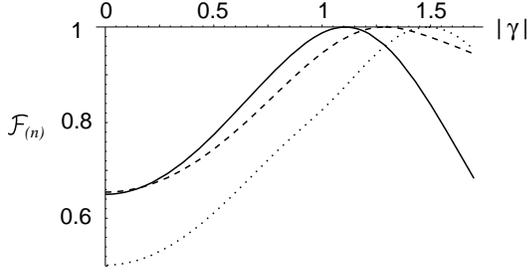}}}
\caption{The fidelity ${\cal F}_{(n)}$ between the ideal SCSs $W^\pm_{\rm scs}(\alpha)$
with amplitude $|\gamma|$ and
the output states $W^{(n)}_{\rm out}(\alpha;X^+_{\rm r}=0)$ from the
$n$-photon input Fock states for $n=2$ (solid line),
$n=3$ (dashed line), and $n=4$ (dotted line).
Note that the target state is the ideal even (odd) SCS
when $n$ is an even (odd) number.  
}
\label{fid}
\end{figure}

Our calculation can be extended for the cases of
$n=3$ and $n=4$. 
We have obtained the fidelity
between the output state $W^{(3)}_{\rm out}(\alpha)$
and the ideal odd SCS
for $X^+_{\rm r}=0$ as
\begin{equation}
{\cal F}_{(3)}(X^+_{\rm r}=0)=\frac{4 \gamma^2e^{2s} A K^\frac{7}{2}
L^2(\coth\gamma^2-1)}
{3 (3+2e^{4s})(1+3e^{2s})^7}
\label{f2}
\end{equation}
where $L=3+12e^{2s}+(9-8\gamma^2)e^{4s}$.
The even state is the target state for $n=4$, and in this case
the fidelity for $X^+_{\rm r}=0$ is 
\begin{equation}
\label{f3}
{\cal F}_{(4)}(X^+_{\rm r}=0)=\frac{ A K^\frac{9}{2}
M^2}
{3 (1+e^{2\gamma^2})(1+3e^{2s})^9(3+24e^{4s}+8e^{8s})}
\end{equation}
with
$M=3+24e^{2s}+(66-48\gamma^2)e^{4s}-24(8\gamma^2-3)e^{6s}
+(64\gamma^4-144\gamma^2+27)e^{8s}$.
The fidelities (\ref{f1}), (\ref{f2}) and (\ref{f3})
for the maximized squeezed parameter, i.e., $\max_{s} {\cal F}_{(n)}(X=0)$,
have been plotted for amplitude $\gamma$ of the SCSs in Fig.~\ref{fid}.
It shows that the fidelity becomes close to one,
for a given number $n$,
when the squeezed parameter and the amplitude are 
properly optimized. One can find
the amplitude of the cat state that maximizes
the fidelity for a given number $n$. 
The maximum fidelity ${\cal F}_{(2)}=0.9997$ is obtained when 
$s=-0.37$ ($2.95$~dB) 
and $\alpha=1.1 i$. 
The same result ${\cal F}_{(3)}=0.9999$ (${\cal F}_{(4)}=0.9997$)
 can be obtained for $n=3$ ($n=4$) when $s=-0.34$ 
 ($s=-0.37$)  and $\alpha=1.29 i$.
($\alpha=1.49 i$).

\begin{figure}
\centerline{\scalebox{0.6}{\includegraphics{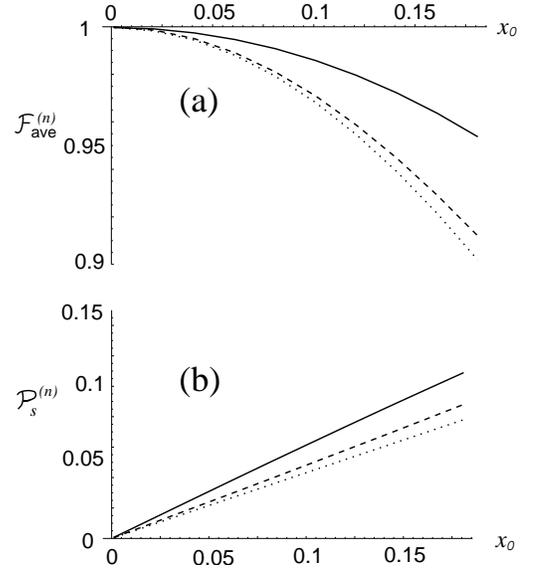}}}
\caption{
(a) 
The average fidelity ${\cal F}^{(n)}_{\rm ave}$
between the ideal SCS and the output state 
of the input Fock state $|n\rangle$
and (b) the success probability ${\cal P}^{(n)}_{\rm s}$
for varying threshold $x_0$.
The beam splitter reflectivity is $R=1/2$.
{\it Solid line} - $n=2$,
the ancillar squeezing is $s=-0.37$, and
the amplitude of the target SCS is $|\gamma|=1.1$.
{\it Dashed line} - $n=3$,
the ancillar squeezing is $s=-0.34$, and
the amplitude of the target SCS is $|\gamma|=1.29$.
{\it Dotted line} - $n=4$,
the ancillar squeezing is $s=-0.37$, and
the amplitude of the target SCS is $|\gamma|=1.49$.
}
\label{fig:cat}
\end{figure}

\begin{figure}[ht!]
\centerline{(a)}
\centerline{\scalebox{0.43}{\includegraphics{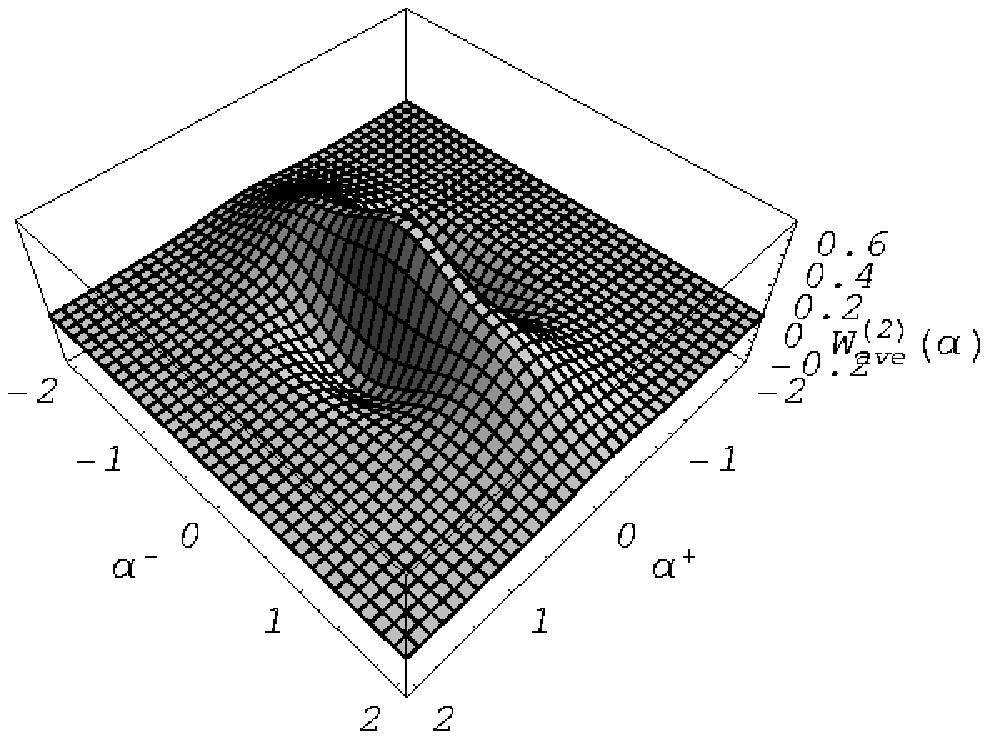}}
\scalebox{0.43}{\includegraphics{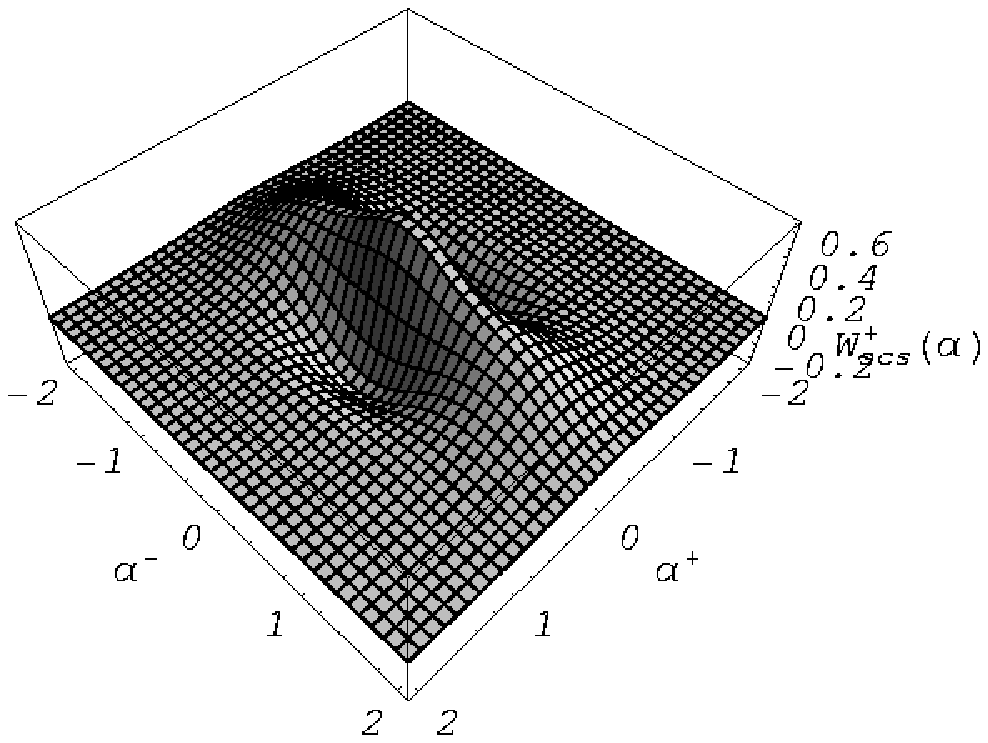}}}
\centerline{(b)}
\centerline{\scalebox{0.43}{\includegraphics{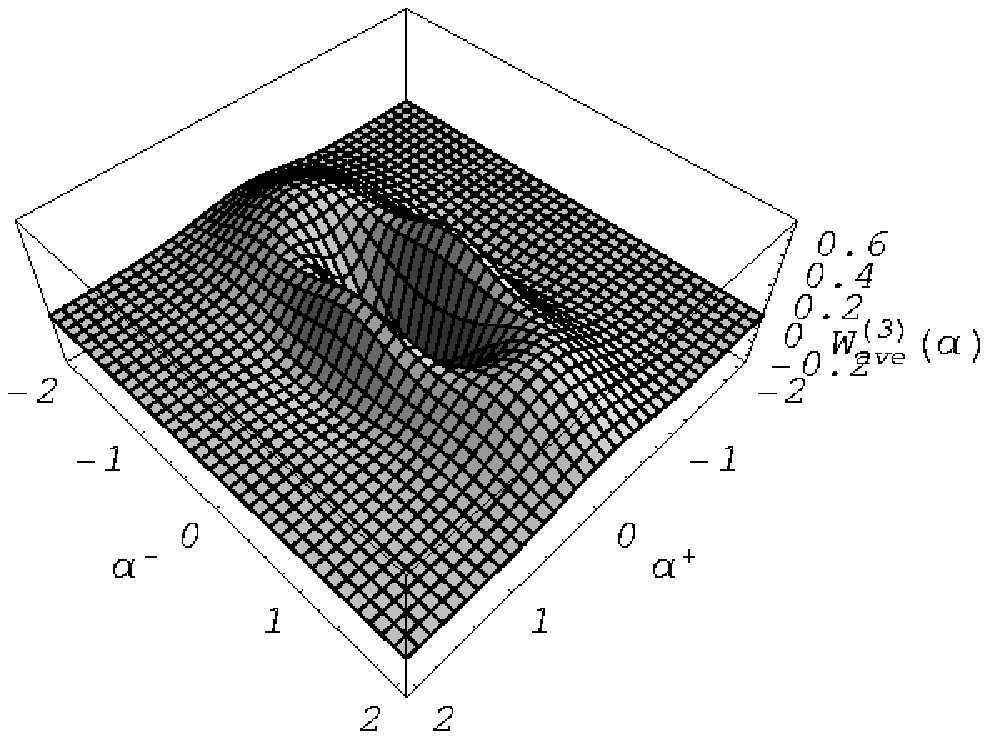}}
\scalebox{0.43}{\includegraphics{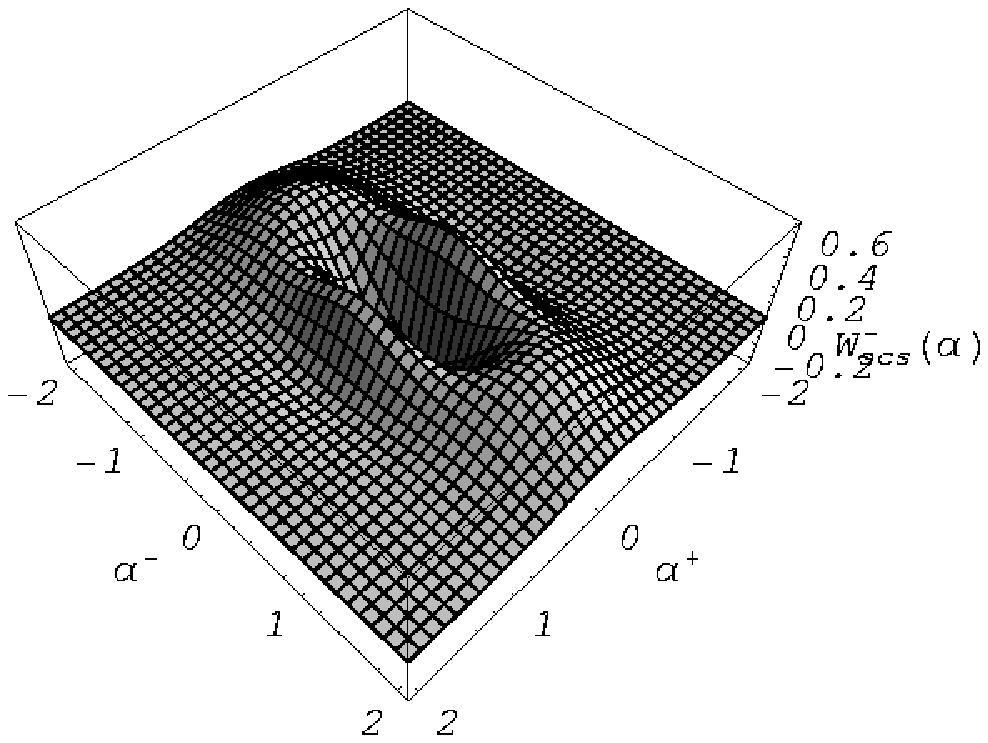}}}
\centerline{(c)}
\centerline{\scalebox{0.43}{\includegraphics{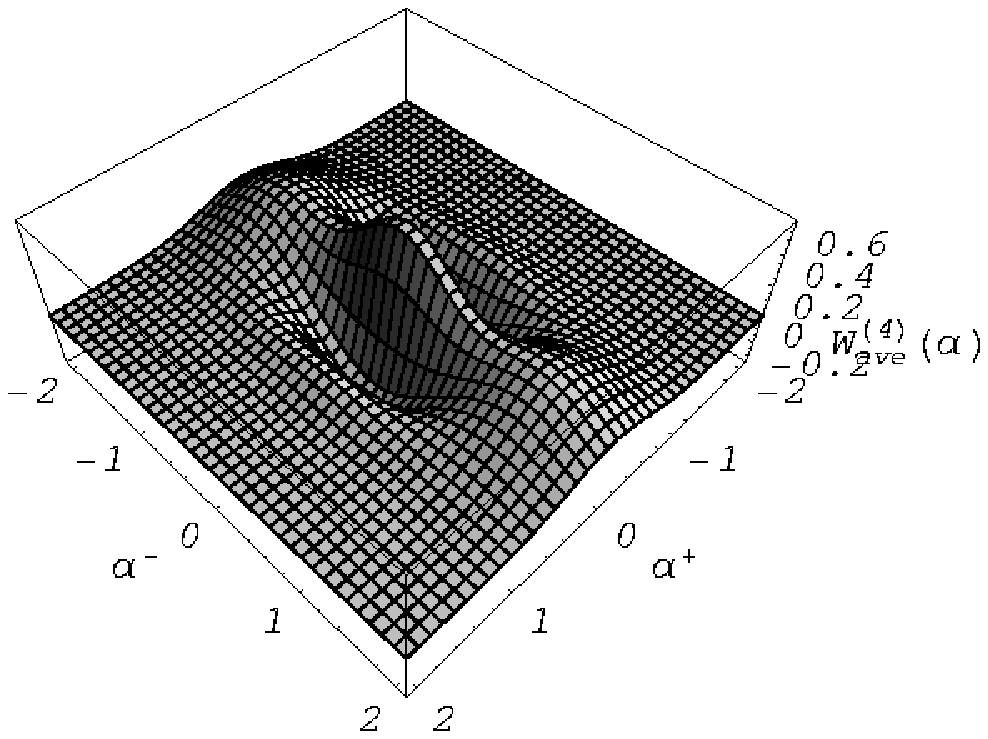}}
\scalebox{0.43}{\includegraphics{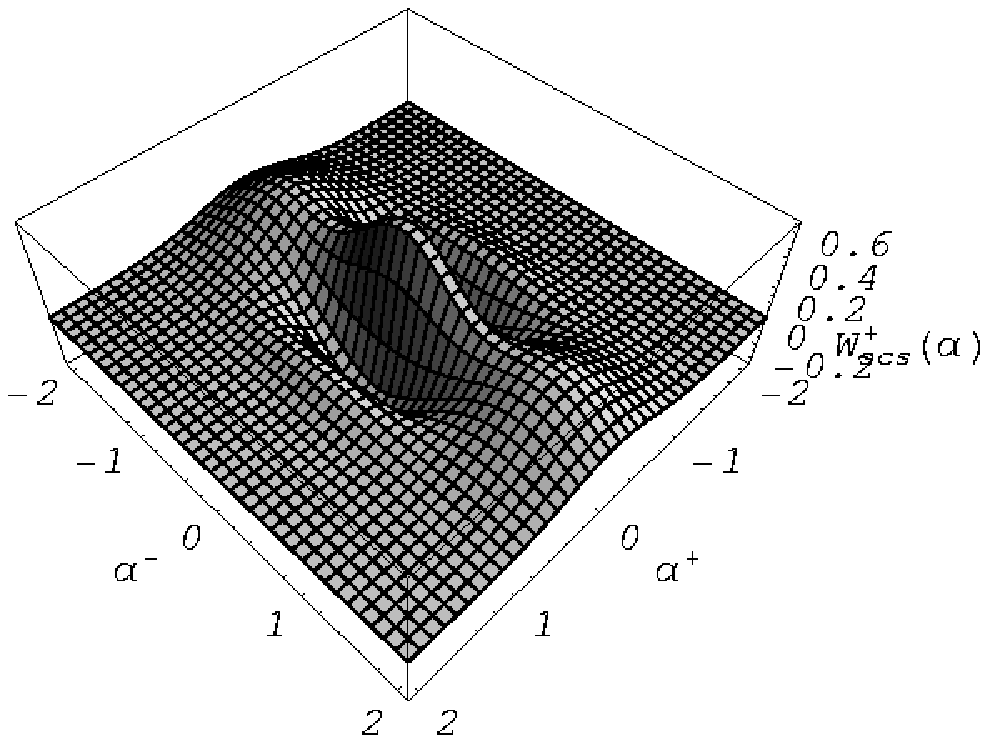}}}
\caption{
The average Wigner functions 
$W^{(n)}_{\rm ave}$ of the output state
corresponding to ${\cal F}^{(n)}_{\rm ave}=0.99$
for (a) $n=2$,
$x_0=0.084$ and ${\cal P}^{(2)}_{\rm s}=0.055$,
for (b) $n=3$, $x_0=0.0583$ and ${\cal P}^{(3)}_{\rm s}=0.0309$,
and for (c) $n=4$, $x_0=0.0555$ and ${\cal P}^{(4)}_{\rm s}=0.0254$.
It is clearly observed that the average output Wigner functions (left hand side)
are virtually identical to those of ideal even ($n=2,4$) and odd ($n=3$) SCSs
(right hand side) with the corresponding amplitudes.}
\label{2photon_Input}
\end{figure}

We have calculated the success probability ${\cal P}^{(n)}_{\rm s}(x_0)$
and the average fidelity
${\cal F}_{\rm ave}^{(n)}$ for $n=2$ to $n=4$ as described above 
and plotted  for varying threshold $x_0$ in Fig.~\ref{fig:cat}.
Our numerical calculations show that
${\cal F}_{\rm ave}^{(2)}=0.999$ is obtained
for 1.4\% when $X_0=0.022$ for the two-photon input state,
and ${\cal F}_{\rm ave}^{(3)}=0.999$
(${\cal F}_{\rm ave}^{(4)}=0.999$ )
is obtained for 0.8\% ($0.6\%$) when $X_0=0.017$ ($X_0=0.014$).
The success probability is improved if the required fidelity
is ${\cal F}_{\rm ave}^{(n)}=0.99$.
Using the same conditions,
 ${\cal F}_{\rm ave}^{(2)}=0.99$ is obtained
for 5.2\% when $X_0=0.084$. The same fidelity ${\cal F}_{\rm ave}^{(3)}=0.99$
(${\cal F}_{\rm ave}^{(4)}= 0.99$)
is obtained for 2.8\% ($2.4\%$) when $X_0=0.058$ ($X_0=0.055$).
In Fig.~\ref{2photon_Input}, the average output Wigner functions 
$W^{(n)}_{\rm ave}$ for the input $n$-photon Fock states
look identical to the Wigner functions of the
corresponding even and odd SCSs due to the high fidelities ($=0.99$).

Based on our results, we can conjecture that an $n$-photon Fock state
can be converted to a SCS using our scheme, and  
the parity of the SCS is determined by
the parity of the input Fock state.
It will be interesting
to prove our conjecture for an arbitrary number $n$,
which is yet beyond the scope of our paper.

\subsection{Squeezing coherent states}

We now consider the case of a Gaussian state, i.e.,
an {\it unknown} coherent state, $|\gamma\rangle$, as the input. 
The post-selection scheme for $X^+_{\rm r}=0$ transforms the coherent state as 
\begin{equation}
D(\gamma)|0\rangle\longrightarrow 
D\big(\sqrt{T}[e^{2s^\prime}
\gamma^{+}+i \gamma^{-} ]\big)S(s^\prime)|0\rangle,
\label{eq:ppt}
\end{equation}
where $D(\gamma)={\rm exp}[\gamma\hat{a}^\dagger-\gamma^{*}\hat{a}]$ 
is the displacement operator and the output squeezing $s^\prime$ is
\begin{equation}
s^\prime=-\frac{1}{2}\log[T+e^{-2s}R].
\label{equ:sp2}
\end{equation}
with the ancillar squeezing $s$.
These relations can be obtained by analyzing the input and output
Wigner functions.  
Note that Eq.~(\ref{equ:sp2}) for coherent state inputs is
identical to Eq.~(\ref{equ:sp1}) for single photon inputs.
The transform in Eq.~(\ref{eq:ppt}), illustrated in Fig.~\ref {ExptSetup0}~(b),
has an interesting property in that it preserves the purity 
of the input coherent state, i.e. the output is a 
minimum uncertainty state, independent of the amount 
of squeezing of the ancilla state.
In the ideal limit of perfect phase squeezing of the ancilla state
 $s\!\rightarrow\infty\!$, 
the post-selection scheme works as an ideal 
single-mode squeezer for arbitrary input coherent states
$D(\gamma)|0\rangle\rightarrow S(s^\prime)D(\gamma)|0\rangle$.
In this case, the output squeezing is 
$s^\prime\rightarrow-\ln [T]/2$.
On the other hand, if $s<0$, $s^\prime$ is not limited by the beam splitter ratio
while the transformation does not work as an ideal squeezer in the limit of
the infinite squeezing.

\begin{figure}
\centerline{\scalebox{0.5}{\includegraphics{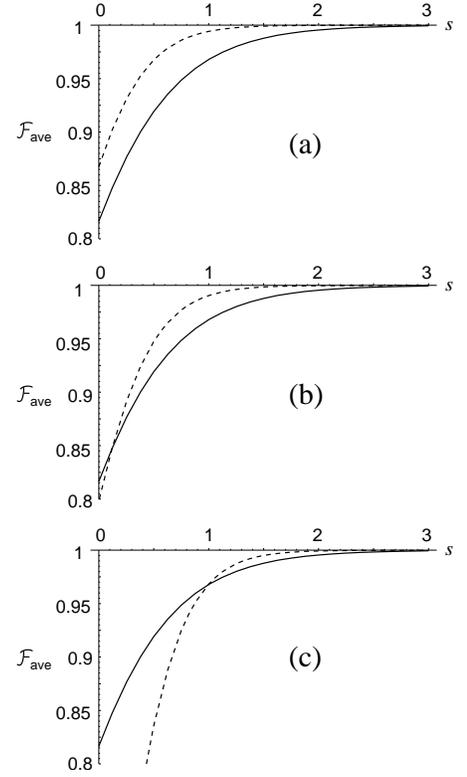}}}
\caption{The average fidelity ${\cal F}_{\rm ave}$
between the output state from a coherent state input $|\gamma\rangle$ 
and an ideally squeezed coherent state by
the feed-forward scheme (solid line) and
the post-selection scheme (dashed lime)
against the ancillar squeezing $s$ of
the resource squeezed vacuum.
$R=1/2$, $|X|<0.025$, and $s^\prime = 0.346574$.
The coherent amplitudes  are (a) $|\gamma|=0.5$,
(b) $|\gamma|=1$ and (c) $|\gamma|=2$.
The threshold for the postselection scheme is $x_0=0.025$.
} 
\label{fig:smalla}
\end{figure}

It is interesting to note that the transformation 
(\ref{eq:ppt}) using the postselection scheme
outperforms the feedforward scheme with the standard gain \cite{Fil05}
to squeeze unknown coherent states
for small amplitudes as shown in Fig.~\ref{fig:smalla}.
The standard gain for the feedforward scheme is
\begin{equation}
g=\sqrt{\frac{R}{T}}
\end{equation}
which makes the center of the average output state the same 
as it of the input state.  
Fig.~\ref{fig:smalla} also implies that 
when the ancillar squeezing becomes larger, the postselection scheme
outperforms the feedforward scheme for the larger area of $\gamma$.
However, it can be shown
for a coherent state input that the feed-forward scheme also
works as the purity preserving transform in Eq.~(\ref{eq:ppt})
when the electronic gain is set to be
\begin{equation}
g=
\frac{(1-e^{-2s})
\sqrt{RT}}
{(e^{-2s}R+T)}.
\end{equation}
Therefore, the post-selection scheme can perform interesting tasks
which cannot be achieved by the feed-forward scheme {\it for non-Gaussian inputs},
while its transformation {\it for Gaussian inputs} can be achieved by 
a modification of the feedforward scheme \cite{JF05}.

\section{Experimental demonstration for coherent state inputs} 

As shown in the theoretical section, the post-selection 
protocol is highly efficient for transforming both 
non-Gaussian input Fock states and Gaussian input 
coherent states. We experimentally demonstrated 
the principle of the post-selection protocol using 
displaced coherent input states with realizable 
ancilla state squeezing. We characterized the efficacy of the 
post-selection protocol by measuring the fidelity of 
post-selected output state 
compared to the target state which is the
ideal squeezed transform of the input state in the case of 
perfect ancillary state squeezing.  
For the experiment, the quantum states we considered
reside at the sideband frequency $(\omega)$ of the 
electromagnetic field. Without the loss of generality,
we denote the quadratures of these quantum states as 
$\hat{X}^{\pm}=\langle\hat{X}^{\pm}\rangle+\delta\hat{X}^{\pm}$, 
where $\langle\hat{X}^{\pm}\rangle$ are the quadrature 
expectation values, and where the quadrature variances 
are expressed by 
$V^{\pm}=\langle(\delta\hat{X}^{\pm})^{2}\rangle$. 

\begin{figure}[ht!]
\begin{center}
\includegraphics[width=\columnwidth]{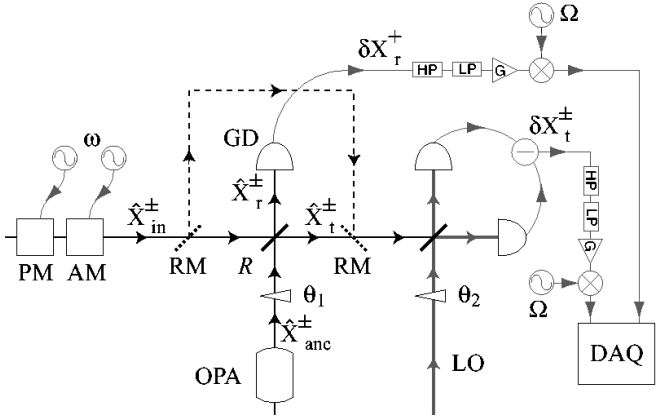}
\caption{(a) Schematic of the post-selection protocol. 
$\hat{X}_{\rm in}^{\pm}$: amplitude (+) and phase (-) 
quadratures of the input state; (anc) 
ancilla state; (r) reflected; (t) transmitted;  and (out)
post-selected output state. 
$R$: beam-splitter reflectivity; GD: gate detector;
RM: removable mirror; OPA; optical parametric 
amplifier; LO: local oscillator; AM/PM:
amplitude/phase modulator; 
$\theta_{1}$/$\theta_{2}$: optical phase 
delay; HP/LP:
high/low pass filters; G: radio frequency amplifier; $\omega$: 
frequency of displaced coherent state at 6.81~MHz; 
$\Omega$: electronic local oscillator at 6.875~MHz; 
DAQ: data acquisition system.
}\label{ExptSetup}
\end{center}
\end{figure}

Fig.~\ref{ExptSetup} shows the experimental setup
for the post-selection protocol. 
We used a hemilithic cavity MgO:LiNbO$_3$ 
below-threshold optical parametric amplifier, to
produce an amplitude squeezed field at 1064~nm
with a squeezing coefficient of  $s=0.52\pm 0.03$, 
corresponding to quadrature variances of 
$V^{+}_{\rm anc}=-4.5\pm 0.2$~dB and 
$V^{-}_{\rm anc}=+8.5\pm 0.1$~dB with respect to the 
quantum noise limit. The amplitude 
squeezed state was a slightly mixed
state due to decoherence as a result of
losses in the optical parametric amplifier. 
More detail of this experimental production 
of squeezing is given in~\cite{Bow03}. 
The displaced coherent states were produced at the 
sideband frequency of 6.81~MHz of the
1064~nm laser field using standard amplitude 
and phase electro-optic modulation techniques 
of the laser field~\cite{Bow03}. 
To produce the phase squeezed ancillary state 
required in the protocol, the amplitude squeezed ancilla 
state $\hat{X}^{\pm}_{\rm anc}$ was transformed
to a phase squeezed state by interfering it with 
the displaced input coherent state $\hat{X}^{\pm}_{\rm in}$, 
which had a much larger coherent amplitude,  
on the beam-splitter (with reflectivity $R$) with a relative 
optical phase shift of $\pi/2$. This optical interference 
yielded two output states that were phase squeezed. 
The optical fringe visibility between the two fields was 
$\eta_{\rm vis}=0.96\pm 0.01$.

We directly detected the amplitude quadrature of the 
reflected state $\hat{X}^{+}_{\rm r}$ 
using a detector, denoted by the gate-detector in
Fig.~\ref{ExptSetup} , which
had a quantum efficiency of $\eta_{\rm det}=0.92$ and 
an electronic noise of $6.5$~dB below the 
quantum noise limit. The post-selection process 
could in principle be 
achieved using an all optical setup, using an 
optical switch for example, but here
we post-selected {\it a posteriori} the  
quadrature measurements of the transmitted state, 
$\hat{X}^{\pm}_{\rm tran}$, 
which were measured using a balanced homodyne 
detector. The total homodyne detector efficiency was
$\eta_{\rm hom}=0.89$, with the electronic noise
of each detector $8.5$~dB below the quantum noise limit. 
To characterize the protocol, for each 
experimental run we also measured 
the quadratures of the input displaced coherent state, 
$\hat{X}^{\pm}_{\rm in}$, using the 
same homodyne detector via a pair of removable 
mirrors (Fig.~\ref{ExptSetup}). 
To ensure accurate results, the total 
homodyne detector inefficiency was inferred out 
of all quadrature measurements for the 
post-selected and input quantum states~\cite{Bow03}. 

The electronic photocurrents of the detected quantum 
states (at a sideband frequency of 6.81~MHz) from the gate 
and homodyne detectors were electronically 
filtered, amplified and demodulated down to 25~kHz
using an electronic local oscillator at 6.785~MHz. 
The resulting photocurrents were 
digitally recorded using a NI PXI 5112 data acquisition 
system operating at a sample rate of 100~kS/s. Typically, 
$5\times 10^{5}$ ($2\times 10^{6}$)
samples of data were taken for the $\hat{X}^{\pm}_{\rm in}$  
($\hat{X}^{\pm}_{\rm t}$ and 
$\hat{X}^{\pm}_{\rm r}$) quadrature measurements, whilst
$5\times 10^{5}$ samples of data were 
taken for the noise calibration data. We used computational 
methods to apply a 25~kHz band-pass filter to the data
centered at 25~kHz, removing technical noise at 
0~Hz, and ensuring that the resulting 
frequency noise spectrum was homogenous. 
The data was demodulated to 0~Hz using a 
digital local oscillator at  25~kHz,  
and down-sampled to a sample
rate of 25~kHz, so that it could be 
directly analyzed 
in the temporal domain. 

We post-selected the quadrature measurements 
of the transmitted state, $\hat{X}^{\pm}_{\rm t}$, 
which satisfied the 
threshold criteria $|X^{+}_{\rm r}|<x_0$. 
The post-selection threshold was dependent on the
beam-splitter reflectivity used in the protocol. 
For a beam-splitter reflectivity of  
$R=0.75$ and $R=0.5$ we used an 
experimentally optimized threshold of  
$x_0= 0.009$ and $x_0= 0.005$ respectively.
Hence, the post-selection threshold was 
independent of the input state, but dependent on the 
beam-splitter reflectivity. 

\begin{figure}[ht!]
\begin{center}
\includegraphics[width=\columnwidth]{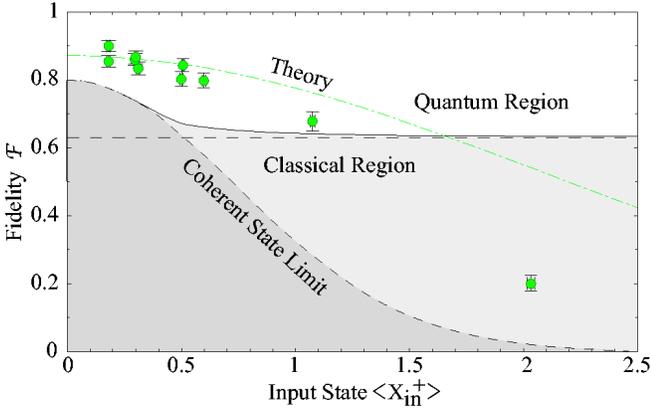}
\caption{Experimental fidelity of the post-selected output state
for varying amplitude quadrature expectation values 
of the input coherent state $|\gamma^{+}|\equiv
\langle\hat{X}^{+}_{\rm in}\rangle$, for $R=0.75$ and
$x_0= 0.009$ Dark grey region: classical fidelity limit for an
ancilla vacuum state; Light grey region: classical fidelity limit.
Dot-dashed line: calculated theoretical curve including 
experimental losses and the finite post-selection threshold.
}\label{Fidelity75pcent}
\end{center}
\end{figure}
\begin{figure}[ht!]
\begin{center}
\includegraphics[width=\columnwidth]{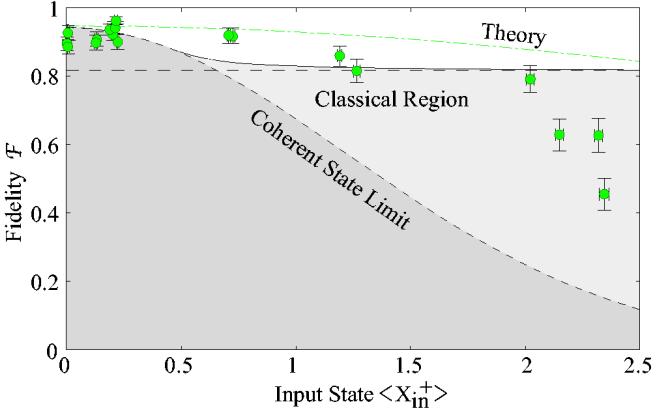}
\caption{Experimental fidelity of the 
post-selected output state for varying 
amplitudes $\langle\hat{X}^{+}_{\rm in}\rangle$ of the input
coherent state, for $R=0.75$ and
$x_0= 0.009$. 
}\label{Fidelity50pcent}
\end{center}
\end{figure}

We characterized the efficacy of the post-selection  
 protocol as an ideal single mode squeezer, by
determining the fidelity of the post-selected output state 
with a target state that is an ideal squeezed operation 
of the input state [Eq.~(\ref{eq:ppt})]. The Wigner function of the 
ideal squeezed input state is given by 
$W_{\rm out}(\gamma;s^\prime)$, where 
$s\rightarrow\infty$ and $s^\prime\rightarrow-\ln [T]/2$. 
In this case, the fidelity is given by ${\cal F}
(X^+_{\rm r})\pi\int^\infty_{-\infty} d^2\gamma W_{\rm expt}(\gamma;X^+_{\rm r})
W_{\rm out}(\gamma;s^\prime)$, where 
$W_{\rm expt}(\gamma;X^+_{\rm r})$ is the 
Wigner function of the post-selected output state. 
From this expression the average fidelity 
${\cal F}_{\rm ave}$ for a post-selection threshold 
$x_0$ can be calculated. 
This corresponds to unity fidelity 
$\mathcal{F}_{\rm ave}=1$ in the limit of ideal ancilla state 
squeezing and $X^+_{\rm r}=0$.
 In the experiment, the input state was not coherent, 
but rather a slightly mixed state due to inherent low-frequency 
classical noise on the laser beam, with
quadrature variances of $V^{+}_{\rm in}=1.13\pm 0.02$ 
and $V^{-}_{\rm in}=1.05\pm 0.02$, with respect to the 
quantum noise limit. 
Hence, we calculated the fidelity with respect to the 
ideal squeezing transform of the experimental input 
state. Fig.~\ref{Fidelity75pcent} shows the classical fidelity limit, 
in the case where the ancilla squeezed state is replaced 
with a vacuum state. The classical fidelity limit $\mathcal{F}_{\rm clas}$,
 is maximized by considering additional classical 
 noise on the phase quadrature
of the input vacuum state (Fig.~\ref{Fidelity75pcent}).
Exceeding this classical fidelity limit can only be 
achieved using quantum resources.

Fig.~\ref{Fidelity75pcent} shows the experimental fidelity 
for varying amplitude quadrature expectation values of the 
input states $|\gamma^{+}| \equiv |\langle\hat{X}^+_{\rm in}\rangle|$.
We point out that for the majority of the input states, 
both the amplitude and phase quadrature expectation
values were approximately equal with
$\langle\hat{X}^{+}_{\rm in}\rangle\approx
\langle\hat{X}^{-}_{\rm in}\rangle$. 
For a beam-splitter reflectivity of $R=0.75$, 
we achieved a best fidelity of $\mathcal{F}_{\rm ave}=0.90\pm0.02$ for
an input state $\langle\hat{X}^{+}_{\rm in}\rangle=0.18\pm0.01$, which 
exceeds the maximum classical fidelity of $\mathcal{F}_{\rm clas}=4/5=0.8$.
This post-selected output state had
quadrature variances of $V^{+}_{\rm out}=4.70\pm0.11$ 
and $V^{-}_{\rm out}=0.51\pm0.01$. The mean quadrature 
displacement gains, $g^{\pm}=\langle\hat{X}^{\pm}_{\rm out}\rangle/
\langle\hat{X}^{\pm}_{\rm in}\rangle$,
which measure the ratio of the quadrature expectation values 
of the post-selected output state with respect to the 
input state, were measured to be
$g^{+}=0.71\pm0.16$ and  $g^{-}=0.50\pm0.06$. This is
compared with the ideal case of perfect ancilla 
state squeezing and a post-selection threshold of 
$\hat{X}^{+}_{\rm r}=0$, where the ideal theoretical gains are 
$g^{+}_{\rm ideal}=2$ and $g^{-}_{\rm ideal}=1/2$. 
The phase gain was controlled by 
the beam-splitter transmittivity, whilst 
the amplitude gain was less than the ideal case  
due to finite ancilla state squeezing, 
finite post-selection threshold
and experimental losses. 
The quantum nature of the post-selection 
protocol is demonstrated by the experimental fidelity 
results that exceed the classical fidelity limit in 
Fig.~\ref{Fidelity75pcent}.
For large input states $\langle\hat{X}^{+}_{\rm in}\rangle$, 
the experimental fidelity was less than the
theoretical prediction due to
electronic detector noise and the finite 
resolution of the data acquisition 
system, resulting in a smaller 
post-selected output state 
$\langle\hat{X}^{+}_{\rm in}\rangle$
and a corresponding 
decrease in the experimental fidelity.

Fig.~\ref{FidelityVarianceGain}~(a) illustrates how
the experimental fidelity of a post-selected state 
transitions from the classical to
 the quantum fidelity region by 
decreasing the post-selection threshold 
and corresponding probability of success. 
Fig.~\ref{FidelityVarianceGain}~(b) shows how 
the corresponding mean amplitude quadrature 
displacement gain $g^{+}$ increases as a result of this 
process, whilst  the phase gain 
remains approximately unchanged. 
Similarly, Fig.~\ref{FidelityVarianceGain}~(c) shows how the 
amplitude quadrature variance 
of the post-selected state $V^{+}_{\rm out}$
is reduced, whilst the phase quadrature 
variance $V^{-}_{\rm out}$ remains approximately unchanged, 
as the post-selection threshold is decreased

\begin{figure}[ht!]
\begin{center}
\includegraphics[width=\columnwidth]{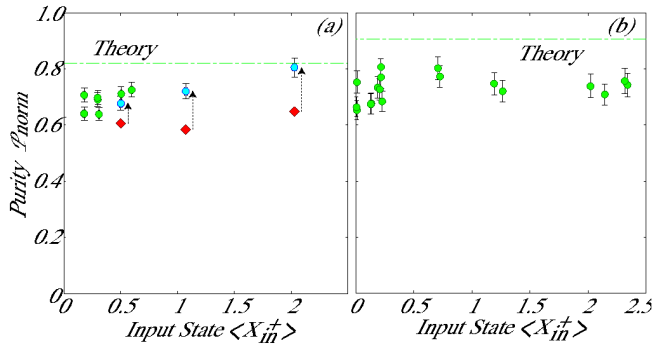}
\caption{ (a) Experimental normalized purity for 
varying input states 
$\langle\hat{X}^{+}_{\rm in}\rangle$, 
for $R=0.75$ and $x_0= 0.009$. 
Dash arrows show purity 
prior to (diamonds) and 
after (circles) post-selection. 
Dot-dashed line: calculated theoretical 
prediction of the experiment.
(b) Experimental purity for varying input state
$\langle\hat{X}^{+}_{\rm in}\rangle$, for $R=0.5$ and 
$x_0= 0.005$. 
}\label{ExptPurity}
\end{center}
\end{figure}

We also characterized the experiment in terms of 
the purity of the post-selected output state, defined 
as $\mathcal{P}={\rm tr}(\rho^{2}_{\rm out})$. In the case of 
Gaussian states, the purity of the output state can be expressed as
$\mathcal{P}=(V^{+}_{\rm out}V^{-}_{\rm out})^{-1/2}$. 
In the ideal case of a lossless experiment and a 
post-selection threshold $X^+_{\rm r}=0$, the  protocol is a 
purity preserving transform, independent of the input state 
and the amount of squeezing of the ancilla state. 
In the experiment, as the input states were slightly mixed, 
\begin{widetext}
\begin{center}
\begin{figure}[ht!]
\includegraphics[width=\columnwidth]{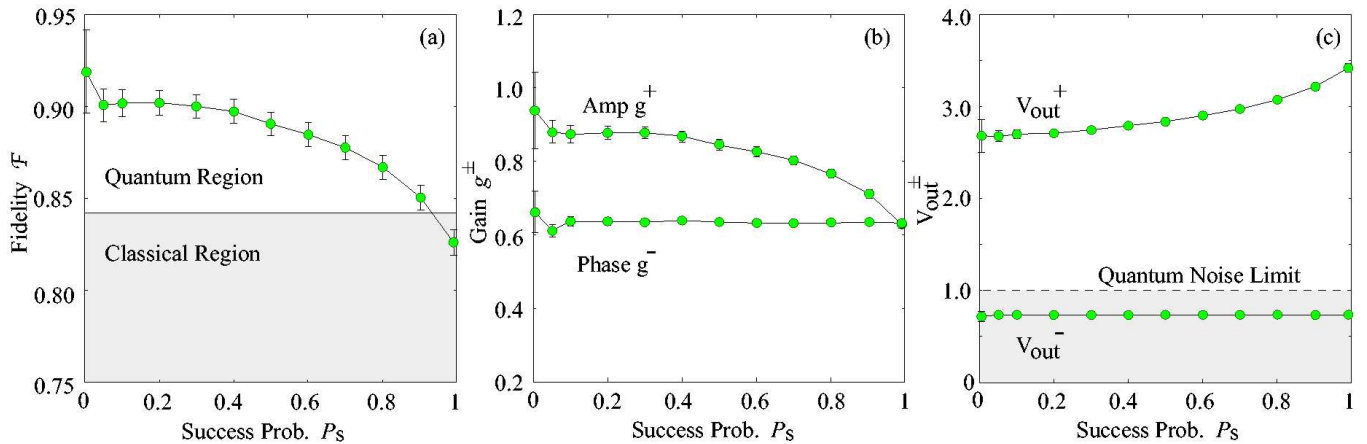}
\caption{(a) Experimental fidelity of the post-selected 
output state for varying success probability corresponding to 
varying post-selection threshold. $R=0.5$ and $\langle\hat{X}^{+}_{\rm in}\rangle=0.71$
Light grey region: classical fidelity region. 
(b) Experimental optical quadrature gains 
$g^{\pm}=\langle\hat{X}^{\pm}_{\rm out}\rangle/
\langle\hat{X}^{\pm}_{\rm in}\rangle$
for varying success probability 
(c) Experimental quadrature variances  
of the post-selected output state
$V^{\pm}_{\rm out}$
for varying success probability.
}\label{FidelityVarianceGain}
\end{figure}
\end{center}
\end{widetext}
we calculated the purity of the post-selected output state, 
normalized to the purity of the input state, which is given by 
$\mathcal{P}_{\rm norm}(V^{+}_{\rm out}V^{-}_{\rm out})^{-1/2}/(V^{+}_{\rm in}V^{-}_{\rm in})^{-1/2}$.
Fig.~\ref{ExptPurity}~(a) shows the experimental purity 
of the post-selected output state for varying input states, 
which illustrates how the purity is improved via the 
post-selection process.
For a beam-splitter reflectivity of $R=0.75$ we achieved a best 
purity of $\mathcal{P}_{\rm norm}=0.81\pm0.04$ 
for an input state of $\langle\hat{X}^{+}_{\rm in}\rangle=2.03\pm0.02$.
Fig.~\ref{ExptPurity}~(a) shows that  
the purity of the post-selected output states 
were approximately independent of the input states,
for a large range of input states.

We also implemented the post-selection scheme  
for a beam-splitter reflectivity of $R=0.5$. In this case,
we measured a best fidelity of $\mathcal{F}_{\rm ave}=0.96\pm0.01$, 
which exceeded the maximum classical fidelity of 
$\mathcal{F}_{\rm clas}=\sqrt{8}/3\approx0.94$ as 
shown in Fig.~\ref{Fidelity50pcent}. We measured 
a best experimental normalized purity of 
$\mathcal{P}_{\rm norm}=0.80\pm0.04$ which is shown
in Fig.~\ref{ExptPurity}~(b).

\section{Remarks}

We have investigated a continuous-variable 
conditioning scheme based on a beam-splitter interaction, 
homodyne detection and an ancilla squeezed vacuum state \cite{Lance06}. 
It transforms input Fock states to
squeezed single photons and SCSs,
which have applications in the field of quantum information,
with realizable squeezing of the ancilla state
\cite{Lance06}.
We have found that a SCS with well defined parity and 
high fidelity can be generated from a Fock state of $n\leq4$,
and we conjecture that this can be generalized for an arbitrary $n$.
Further, for 
Gaussian coherent states, this technique 
provides an alternative to continuous electro-optic 
feed-forward schemes.
The postselection scheme provides
an interesting transformation for coherent states which results in 
higher fidelities to the ideal squeezing operation
when amplitudes are small.
All the interesting 
features of the post-selection scheme for {\it non-Gaussian} inputs 
discussed in this paper {\it cannot} be achieved using
the feed-forward method in Ref.~\cite{Fil05}.
We have described, in detail, the experimental demonstration
of the principles of this 
scheme using coherent states, where 
fidelities were measured that
are only achievable using quantum resources \cite{Lance06}.

We thank the Australian Research Council for financial 
support through the Discovery Program.

\end{document}